# Improved measurement results for the Avogadro constant using a $^{28}$Si-enriched crystal

Y Azuma[1], P Barat[2], G Bartl[3], H Bettin[3*], M Borys[3], I Busch[3], L Cibik[3], G D'Agostino[4], K Fujii[1], H Fujimoto[1], A Hioki[1], M Krumrey[3], U Kuetgens[3], N Kuramoto[1], G Mana[4], E Massa[4], R Meeß[3], S Mizushima[1], T Narukawa[1], A Nicolaus[3], A Pramann[3], S A Rabb[5], O Rienitz[3], C Sasso[4], M Stock[2], R D Vocke Jr[5], A Waseda[1], S Wundrack[3] and S Zakel[3]

[1] National Metrology Institute of Japan NMIJ, 1-1-1 Umezono, Tsukuba, Ibaraki 305-8563, Japan
[2] Bureau International des Poids et Mesures BIPM, Pavillon de Breteuil, 92312 Sèvres Cedex, France
[3] Physikalisch-Technische Bundesanstalt PTB, Bundesallee 100, 38116 Braunschweig, Germany
[4] Istituto Nazionale di Ricerca Metrologica INRIM, str. delle cacce 91, 10135 Torino, Italy
[5] National Institute of Standards and Technology NIST, 100 Bureau Drive, Gaithersburg, MD 20899, USA

*) Corresponding author, e-mail: horst.bettin@ptb.de

**Abstract**

New results are reported from an ongoing international research effort to accurately determine the Avogadro constant by counting the atoms in an isotopically enriched silicon crystal. The surfaces of two $^{28}$Si-enriched spheres were decontaminated and reworked in order to produce an outer surface without metal contamination and improved sphericity. New measurements were then made on these two reconditioned spheres using improved methods and apparatuses. When combined with other recently refined parameter measurements, the Avogadro constant derived from these new results has a value of $N_A = 6.022\ 140\ 76(12) \times 10^{23}$ mol$^{-1}$. The X-ray crystal density method has thus achieved the target relative standard uncertainty of $2.0 \times 10^{-8}$ necessary for the realization of the definition of the new kilogram.

## 1. Introduction

In 2011, the International Avogadro Coordination (IAC) published a comprehensive survey of the different measurements contributing to the determination of the Avogadro constant $N_A$ by counting the atoms in a $^{28}$Si-enriched single crystal [1, 2]. This approach, called the X-ray-crystal-density (XRCD) method, is one of the candidate methods for the realization of the new kilogram definition that is based on fixing the value of the Planck constant $h$. Additionally, an accurate $N_A$ constant is important because its value will be used to define the mole.

One of the principal issues with the 2011 determination of $N_A$ [1] was the fact that the Si spheres used for the volume and mass determinations were covered by a thin layer of metallic contaminant, composed of Ni, Cu and Zn atoms. This contamination probably occurred during the polishing procedure by a contamination of the slurry. The layer was localized in metal silicide islands "floating" on the silicon core matrix [3]. As the optical constants of this layer were unknown, it was later removed by a Freckle$^{TM}$ etch with a selectively high etching rate for silicides [4]. After etching, the sphere AVO28-S8b ('b' for the status after etching) was remeasured at PTB and $N_A$ was redetermined. The consistency of the new result with the formerly obtained $N_A$ value was excellent [5]. Because the etching had degraded the shape of the spheres, they were then reworked at PTB, using a new procedure to improve their sphericity [6]. In June 2013, the surface of the repolished sphere AVO28-S5c ('c' designating the sphere after repolishing) was checked by X-ray fluorescence (XRF) spectrometry, revealing negligible amounts of foreign metals. Moreover, no subsurface damage to the crystal could be detected by high-resolution X-ray diffractometry [7] when compared with strain free etched reference crystal surfaces. The roughness of the surface was below 0.2 nm, near the detection limit of the measurement. The topography of the sphere was measured interferometrically, establishing that the shape of the sphere was defined only by the orientation of the crystallographic axes. Maximum peak-to-valley (p-v) deviations of the diameter were below 70 nm (see section 2.5). The second sphere, AVO28-S8, was then repolished using a slightly different polishing process in order to achieve a better roundness which produced a p-v value below 40 nm for the sphere diameters.

A new measurement campaign was then initiated using these repolished spheres and these new results are presented in this paper together with the refinements to the measurement methods and instrumentation that have occurred since the 2011 review. All of these improvements were necessary to reach a total relative standard uncertainty of $20 \times 10^{-9}$ for $N_A$, an essential goal for the realization of the new kilogram definition.

## 2. Determination of the crystal quantities

The measurement of the Avogadro constant $N_A$, using a silicon crystal, is based upon the following equation,

$$N_A = nM/(\rho a^3), \qquad (1)$$

where $n$ is the number of atoms (8) per unit cell of a silicon crystal and $\rho$, $M$ and $a$ are its density, molar mass and lattice parameter, respectively. Details on counting the atoms in a silicon crystal are given in [1, 2]. A $^{28}$Si-enriched silicon single crystal was grown primarily to reduce the uncertainty of measuring the molar mass $M$. Two 1 kg spheres were manufactured from the crystal, and the density $\rho$ of each sphere was determined from its mass and volume measurements. The sphere surfaces were covered with oxide layers having a total thickness around 2 nm. In order to determine the density of the crystal at the highest levels of accuracy, the surface of each sphere needed to be chemically and physically characterized at an atomic scale so that the density of the crystal could be determined from the mass and volume data, excluding these oxide layers. In this paper, these two parameters are designated as 'core mass' and 'core volume', respectively.

In a real crystal, the lattice spacing and density are affected by impurity atoms and vacancies. For example, interstitial oxygen expands the lattice spacing and increases the unit cell mass, and substitutional carbon shrinks the lattice spacing and decreases the unit cell mass. When the effect of these point defects on the density $\rho$ of the crystal is considered, the simplest way to implement measurement equation (1) is to calculate the mass of an equivalent sphere, having the same core volume and lattice parameter $a$ measured by combined X-ray/optical interferometry, but having Si atoms at all regular sites.

The concentrations of carbon, oxygen, boron and vacancy-related defects have already been reported [1]. The concentration of nitrogen was additionally measured at PTB. INRIM also developed a method based on instrumental neutron activation analysis (INAA) to evaluate the concentrations of various impurity elements. These results are given in section 2.1.

The amount-of-substance fractions of the Si isotopes in the crystal were measured independently by PTB, NMIJ and NIST using isotope dilution and a multicollector inductively coupled plasma mass spectrometer (MC ICP-MS). Instead of using NaOH as solvent and diluent, tetramethylammonium hydroxide (TMAH) was used by these three institutes to reduce the baseline level of the ion current measurements during the mass spectrometry. These results are given in section 2.2.

To measure the lattice parameter, INRIM upgraded a combined X-ray/optical interferometer to further reduce the uncertainty. To demonstrate crystal homogeneity, NMIJ evaluated the crystal perfection using strain topography, carried out by means of a self-referenced X-ray diffractometer at the Photon Factory of the High Energy Accelerator Research Organization (KEK, Japan). Detailed results are given in section 2.3.

Sections 2.4 to 2.6 describe the measurement of the densities of the two $^{28}$Si-enriched spheres. PTB and NMIJ characterized the composition, mass and thickness of the sphere surface layers by XRF, X-ray reflectometry (XRR), and optical spectral ellipsometry (SE). These results are given in section 2.4. The sphere volumes were determined via diameter measurements. NMIJ measured about a thousand diameters for each sphere using an improved optical interferometer with a flat etalon. PTB used a spherical Fizeau interferometer which allowed about $10^5$ diameters to be measured, resulting a complete topographical mapping of the spheres. Details are given in section 2.5. Mass comparisons of the two spheres with Pt–Ir kilogram standards were carried out both in air and under vacuum by the BIPM, NMIJ and PTB. In order to provide a better traceability to the international prototype of the kilogram (IPK), the BIPM revised the mass values for the BIPM calibrations following the Extraordinary Calibration Campaign against the IPK conducted in 2014 [8]. NMIJ and PTB also used the revised mass values of their Pt–Ir kilogram standards that were reported by the BIPM in December 2014. Details are given in section 2.6.

The final $N_A$ values obtained from these measurements are given in section 3 together with their uncertainty budget. In section 4, the XRCD final result for $N_A$ is compared with those from the watt balance experiments.

*2.1. Point defects*

The infrared (IR) absorption measurements of dissolved carbon, oxygen and boron within the silicon crystal have already been reported in [1]. The gradients in the impurity concentrations are caused by the float zone technique used to purify and grow the single crystal. Additionally, the nitrogen present in the AVO28 crystal was determined by infrared measurements using the method developed by Itoh *et al.* [9]. The average content of nitrogen in the spheres amounts to $0.17(10) \times 10^{14}$ cm$^{-3}$ and $1.38(30) \times 10^{14}$ cm$^{-3}$ for the AVO28-S5 and AVO28-S8 spheres, respectively. This yields a mass deficit for the spheres of -0.2(1) µg and -1.4(3) µg, respectively (see section 2.6).

INRIM has also developed a method based on INAA giving direct evidence of the crystal purity with respect to a very large number of elements. Test measurements carried out at the TRIGA Mark II reactor at the University of Pavia (with a thermal neutron flux of $6 \times 10^{12}$ cm$^{-2}$ s$^{-1}$) included fifty-nine elements and achieved a detection limit of less than 1 ng/g for thirty-five elements [10, 11]. Two samples were cut from the AVO28 boule and another purity check is planned in early 2015, using the OPAL reactor of the Australian Nuclear Science and Technology Organisation (with a thermal neutron flux of $20 \times 10^{13}$ cm$^{-2}$ s$^{-1}$). This analysis is expected to include sixty-four elements and should also reach a detection limit of less than 1 ng/g for more than forty elements.

*2.2. Molar Mass*

Following the 2011 report on the status of molar mass measurements for the Avogadro constant [12, 13], a number of additional investigations have been published, providing new molar mass data on additional crystals from the AVO28 boule [14-17]. In addition, a detailed molar mass and amount-of-substance homogeneity study undertaken by the PTB produced data on an additional 14 crystals. The details of this homogeneity study will be published elsewhere [18]; however, the average molar mass result for the 14 new crystals is reported in table 1 while the individual molar masses are reported in table 2. Table 1 also lists all the average molar mass and amount-of-substance fraction results on the AVO28 crystal material published by national metrology institutes (NMIs) to date.

**Table 1.** A summary of the molar mass and amount-of-substance fraction determinations of the AVO28 crystal material. The uncertainties ($k = 1$) in parentheses apply to the last respective digits. Note that measurements prior to 2013 were carried out using solutions of aqueous NaOH while all subsequent measurements used aqueous TMAH.

| NMI | Diluent | $M$/(g/mol) | $x(^{28}\text{Si})$/(mol/mol) | $x(^{29}\text{Si})$/(mol/mol) | $x(^{30}\text{Si})$/(mol/mol) | Ref. |
|---|---|---|---|---|---|---|
| PTB 2011 | NaOH | 27.976 970 27(23) | 0.999 957 50(17) | 0.000 041 21(15) | 0.000 001 29(4) | [12, 13] |
| NRC 2012 | NaOH | 27.976 968 39(24) | 0.999 958 79(19) | 0.000 040 54(14) | 0.000 000 67(6) | [14] |
| PTB 2014 | TMAH | 27.976 970 22(17) | 0.999 957 26(17) | 0.000 041 62(17) | 0.000 001 12(6) | [15] |
| NMIJ 2014 | TMAH | 27.976 970 09(14) | 0.999 957 63(3) | 0.000 041 20(7) | 0.000 001 18(3) | [16] |
| NIST 2014 | TMAH | 27.976 969 880(41) | 0.999 957 701(41) | 0.000 041 223(41) | 0.000 001 076(88) | [17] |
| PTB 2015 | TMAH | 27.976 970 13(12) | 0.999 957 50(12) | 0.000 041 38(12) | 0.000 001 121(14) | [this paper, 18] |

Subsequent to the original development and application of the virtual-element isotope dilution-mass spectrometric ((VE) ID-MS) approach to determining an accurate molar mass of the highly $^{28}$Si-enriched AVO28 boule [19], several studies [20, 21] have examined the analytical problems of Si isotope amount ratio measurements when aqueous sodium hydroxide (NaOH) is used as the diluent or solvent to prepare silicon solutions for inductively coupled plasma mass spectrometry (ICP-MS). The studies all noted that aqueous NaOH could create subtle but significant biases to Si isotope amount ratio measurements. For example, the presence of aqueous NaOH as sample diluent causes ion scattering in the detector region of a MC ICP-MS. This charged background elevates the $^{30}$Si baseline in particular [22]. Normally, this is not a problem when the Si isotope sample signals are more than a few hundred mV.

However, when the sample signals for $^{30}$Si$^+$ and $^{29}$Si$^+$ are very low ($\leq 1$ mV), such elevated baseline signals could lead to an overcorrection of measured sample signals. This is especially problematic for the $^{30}$Si$^+$ sample signal in the AVO28 crystals and could lead to a significant biasing of the $x(^{29}$Si$)/x(^{30}$Si$)$ ratio. This effect is amplified when the NaOH concentration of the matrix solution is increased [22].

Careful analysis of the published data using NaOH as solvent and matrix diluent suggests that this effect, together with possible memory carry-over, were contributing factors that gave rise to the discrepancy between the 2011 molar mass measurements reported by PTB [12, 13] and the 2012 results published by NRC [14] (see table 1). When measuring the AVO28 samples, PTB reported using a mass fraction of 1 mg/g NaOH as sample diluent while NRC reported using 24 mg/g NaOH. The extremely high levels of NaOH in the NRC samples magnified the $^{30}$Si signal of the blank, causing the very low $^{30}$Si signals of the AVO28 material (which were typically $\leq 1$ mV) to be seriously over-corrected. The net effect of this bias was to produce the artificially high $x(^{29}$Si$)/x(^{30}$Si$)$ ratios reported in [14]. A more detailed analysis of the causes of the biasing of the NRC $x(^{29}$Si$)/x(^{30}$Si$)$ ratios is underway [20].

The NRC had also postulated that the PTB samples (and by implication, the NMIJ and NIST samples) had undergone direct contamination by natural silicon [14]. This possibility was ruled out when the absolute mass fraction of silicon in the NaOH material used for the PTB analyses was found to be a factor of ten lower than the concentration necessary to explain the PTB-NRC molar mass differences. Given that these points of contention involve the use of NaOH as solvent and diluent and have not yet been fully resolved, this study has excluded all molar mass data reported using this particular solvent. Instead, the abundant molar mass data acquired using TMAH as solvent and diluent are used to deduce the molar mass results presented in this paper (table 2).

The use of aqueous TMAH solutions, as proposed and carried out for the first time at NIST [17], has many advantages. The extreme enrichment of the $^{28}$Si isotope in the AVO28 material leaves very little $^{29}$Si and almost no $^{30}$Si to be ionized and detected. With TMAH as the matrix diluent, the large flux of Na$^+$ ions was no longer present in the plasma source. This had the positive effect of increasing the intensity of the very small $^{29}$Si$^+$ and $^{30}$Si$^+$ signals by nearly an order of magnitude. Ancillary negative effects, like the ion scattering produced when NaOH was used, were no longer observed. Additionally, the progressive clogging of the skimmer and sampler cone orifices during sample analyses was strongly reduced or absent when using TMAH. This has led to stable silicon ion beam intensities lasting several days or more. The TMAH blanks measured during the analysis of AVO28 materials also showed a more natural silicon isotopic composition when compared with similar NaOH blanks. This attribute also suggests that there has been a real time mitigation of any memory-carryover issues [17]. All of these factors arising from the use of TMAH have served to greatly improve the quality of the molar mass measurements by decreasing the possibility of measurement biasing, particularly from signal suppression causing an overcorrection of the detected $^{30}$Si and $^{29}$Si sample signals.

While only molar mass data taken with TMAH as solvent and diluent are used in this study to compute the average molar mass of the AVO28 boule, an additional blunder check on the accuracy of these measurements was provided by a recent internal study at PTB. A small disc of AVO28 material was analyzed by glow discharge mass spectrometry (GDMS) [21]. This complementary study produced numerical values for the molar mass and the amount-of-substance fractions in agreement with those reported by PTB, NMIJ and NIST within the limits of uncertainty, although the associated uncertainties of this measurement were larger.

INRIM proposed to measure the amount-of-substance fraction of $^{30}$Si by neutron activation [23]. A sample of the AVO28 crystal was analyzed using the TRIGA Mark II reactor. The result, $x(^{30}$Si$) = 0.000\ 001\ 024(18)$ mol/mol [24], further supports the amount-of-substance findings from the PTB, NMIJ and NIST measurements. This measurement is being repeated in Australia using the OPAL reactor.

In 2014, with the specific backing of the IAC, PTB initiated an investigation of the variability of the molar mass and the isotopic composition across the AVO28 boule. The experimental design for this study called for 5 adjacent radial samples (each with a mass of approximately 300 mg) to be taken from three distinct longitudinal positions (parts 4, 7, and 9) of the original crystal ingot (figure 1 in [12]) and analyzed for their molar mass. A de-tailed description of this study will be published elsewhere [18]. However, given that this homogeneity study incorporated most of the state-of-the art improvements in the measurement of Si isotope-amount-ratios developed since 2011, the average and individual molar masses as well as the Si isotope amount fractions derived from these data are presented in tables 1 and 2. The (VE) IDMS approach [19] was used for measuring the molar mass, with the main experimental details given in [13, 15]. All measurements have been made on a commercial MC ICP-MS (Neptune™, Thermo Fisher Scientific). The calibration and sample runs were separated as originally suggested by colleagues from NRC, saving time and material without any significant loss in accuracy. TMAH was used as the solvent and matrix diluent. The marked increase of the $Si^+$ ion signals resulting from the use of TMAH enabled all the Si isotope data to be measured with Faraday detectors ($10^{11}$ Ω resistors). To date, five crystals from part 4, five crystals from part 7 and four crystals from part 9 of the Avogadro boule have been measured (table 2). Because the crystals from parts 4 and 7 bracket part 5 (the source location of sphere AVO28-S5), and parts 7 and 9 bracket part 8 (the source location of sphere AVO28-S8), the new PTB data listed in tables 1 and 2 are fully representative of the spheres AVO28-S5 and AVO28-S8 [18]. When combined together with the data given in [15], these measurements represent the PTB contribution that is combined with the NMIJ and NIST data for the calculation of a new $N_A$.

NMIJ also carried out molar mass measurements of AVO28 crystals using a commercial MC ICP-MS (Neptune™, Thermo Fisher Scientific) with TMAH as the solvent. To correct for the mass bias arising in the plasma and sample source areas, three different blends, as described by NRC [14] were employed. Four AVO28 samples, identified as 5B1.2.2.1, 5B1.2.2.2, 8B3.2.2.1 and 8B3.2.2.5 were analyzed. All of the AVO28 crystals had been cut from the single crystal produced by the IAC specifically to determine the Avogadro constant. The axial positions of the 5B1 series and the 8B3 series are respectively 275 mm and 414 mm from the outer surface of the boule; their radial distance from the center of the $^{28}Si$-enriched crystal ingot is approximately 40 mm. The average molar mass of the four AVO28 crystals was determined to be 27.976 970 09(14) g/mol, with a relative standard uncertainty of $5.2 \times 10^{-9}$ (table 1).

NIST analyzed 4 different crystals of the AVO28 silicon, two proximal to AVO28-S5 (5B2.1.1.3, 5B1.1.1.1) and two taken near AVO28-S8 (8A4.1.1.3, 8B1.1.1.1) [17]. All silicon samples were dissolved and diluted using TMAH and also analyzed on a commercial MC ICP-MS (Neptune™, Thermo Fisher Scientific) in high resolution mode. To correct for measurement mass bias, two independent sets of calibration solutions were created. The AVO28 samples were always run as pairs, with one from part 5 and the other from part 8. These samples were always run together with the calibration solutions. This particular experimental design optimized the detection of any small but measurable heterogeneities in the silicon isotope amount ratios of the different samples. The calibration corrections followed the approach originally developed by PTB [19]. One sample set was measured using only one calibration solution, while the other was analyzed using both solutions. The average molar mass of the four AVO28 crystals measured by NIST was 27.976 969 880(41) g/mol, with a relative standard uncertainty of $1.5 \times 10^{-9}$ [17] (table 1).

In summary, the basis for the average molar mass of the AVO28 boule, with its associated uncertainty, was derived from molar mass measurements using TMAH as solvent and diluent. The individual molar masses of five crystal samples from part 4 (PTB [18, this paper]), one crystal from part 5 (PTB [15]), two crystals from part 5 (NMIJ [16]), two crystals from part 5 (NIST [17]), five crystals from part 7 (PTB [18, this paper]), one crystal from part 8 (PTB [15]), two crystals from part 8 (NMIJ [16]), two crystals from part 8 (NIST [17]) and four crystals from part 9 (PTB [18, this paper]) were used. These data allow a robust molar mass to be calculated for updating the Avogadro constant that is based on 24 different sample crystals spread longitudinally across the AVO28 boule and measured by three different NMIs, each employing a different experimental approach (table 2).

**Table 2.** Compilation of the molar masses measured on 24 individual crystal samples that used TMAH as solvent and diluent (see text for explanation). Uncertainties ($k = 1$) in parentheses apply to the last two digits. Samples are given in the order of their longitudinal position in the original crystal. Sample association with sphere AVO28-S5 or sphere AVO28-S8 is indicated by (S5) and (S8) in the Sample ID column. Note that the 5 crystals from part 7 (PTB-7-1 (S5, S8) to PTB-7-5 (S5, S8)) are included in both calculations of the averages for spheres S5 and S8 because this part lies between the two spheres. The part 7 values have therefore not been arbitrarily associated with one or the other sphere. All molar mass calculations used atomic mass values reported in the AME2012 atomic mass evaluation [25].

| NMI* | Sample ID | $M$ / (g/mol) |
|---|---|---|
| PTB 2015 | PTB-4-1 (S5) | 27.976 970 29(14) |
| PTB 2015 | PTB-4-2 (S5) | 27.976 970 25(13) |
| PTB 2015 | PTB-4-3 (S5) | 27.976 970 00(11) |
| PTB 2015 | PTB-4-4 (S5) | 27.976 970 28(13) |
| PTB 2015 | PTB-4-5 (S5) | 27.976 970 13(13) |
| PTB 2014 | PTB-5 (S5) | 27.976 970 24(17) |
| NIST 2014 | NIST-5-1 (S5) | 27.976 969 842(93) |
| NIST 2014 | NIST-5-2 (S5) | 27.976 970 141(71) |
| NMIJ 2014 | NMIJ-5-1 (S5) | 27.976 970 10(22) |
| NMIJ 2014 | NMIJ-5-2 (S5) | 27.976 970 05(8) |
| PTB 2015 | PTB-7-1 (S5, S8) | 27.976 970 06(12) |
| PTB 2015 | PTB-7-2 (S5, S8) | 27.976 970 09(13) |
| PTB 2015 | PTB-7-3 (S5, S8) | 27.976 969 94(12) |
| PTB 2015 | PTB-7-4 (S5, S8) | 27.976 969 96(12) |
| PTB 2015 | PTB-7-5 (S5, S8) | 27.976 970 00(12) |
| PTB 2014 | PTB-8 (S8) | 27.976 970 20(17) |
| NIST 2014 | NIST-8-1 (S8) | 27.976 969 745(57) |
| NIST 2014 | NIST-8-2 (S8) | 27.976 969 797(90) |
| NMIJ 2014 | NMIJ-8-1 (S8) | 27.976 970 14(21) |
| NMIJ 2014 | NMIJ-8-2 (S8) | 27.976 970 08(21) |
| PTB 2015 | PTB-9-1 (S8) | 27.976 970 08(11) |
| PTB 2015 | PTB-9-2 (S8) | 27.976 970 26(11) |
| PTB 2015 | PTB-9-3 (S8) | 27.976 970 20(11) |
| PTB 2015 | PTB-9-5 (S8) | 27.976 970 33(11) |

*NIST 2014 [17]; NMIJ 2014 [16]; PTB 2014 [15]; PTB 2015 [18, this paper]

The uncertainty weighted mean (UWM) of all 24 results was determined to be 27.976 970 030(38) g/mol with an associated uncertainty expanded by the Birge ratio ($\sigma_B \approx 1.70$) calculated according to [26]. The arithmetic mean including all 24 results was 27.976 970 09 g/mol with a standard deviation of the mean of 0.000 000 03 g/mol. The UWM would normally be the estimator of choice for combining these data, as they show a relatively large spread in their associated uncertainties. However, a data consistency check (chi-squared test) recommended in [27] was carried out, yielding a $X^2_{\text{obs}} \approx 66$ which is larger than the 95$^{\text{th}}$ percentile of $X^2_{0.05,23} \approx 35$, with 23 degrees of freedom. This data set is therefore not entirely internally consistent. This suggests that one or more of the contributions to the overall uncertainty have not been considered fully. Possible sources for the observed inconsistency may come from either the external reproducibility of the measurement due to tiny yet uncontrolled blank variations, the signal detec-tion itself, the reproducibility of the sample preparation, a tiny but detectable variability in the Si isotopic composition, or an as-yet-unknown additional influence. In order to render these data internally consistent, an additional uncertainty contribution with an expectation value of zero was added to the overall uncertainty of the mean, by adapting the recommendations from [28]. The criterion for determining the value of this additional element of uncertainty, $u_{disp}$, is the normalized error [29]. Following [30], $u_{disp}$ was adjusted so that all normalized errors were equal to or less than 1.

Combining the 24 results from PTB, NMIJ, and NIST (table 2), the average molar mass for AVO28 is calculated to be 27.976 970 09(15) g/mol, with an associated relative standard uncertainty of $5.4 \times 10^{-9}$. These data clearly suggest that, at the present level of measurement accuracy and precision, there are no longitudinal or axial molar mass gradients within the AVO28 boule. The individual molar masses of the spheres AVO28-S5 and AVO28-S8 were calculated from the PTB, NMIJ, and NIST results listed in table 2 and labelled "S5" and "S8", respectively. Table 3 summarizes the molar mass data for each separate sphere, as well as the combined data. The average individual molar mass data for the two spheres are indistinguishable, thus the average AVO28 boule value derived by combining all results is the number that is used in the calculation of the new $N_A$ reported in this study.

**Table 3.** Summary of the measured molar masses of spheres AVO28-S5 and AVO28-S8 as well as the average of all molar masses combined. Uncertainties ($k = 1$) in parentheses apply to the last two digits. The final column lists the number of crystals ($n$) used to compute the respective averages. Note that the average molar masses for S5 and S8 each include the 5 molar mass measurements made on the 5 crystals from part 7, as noted in table 2. The average molar mass for the AVO28 boule is the arithmetic mean of all 24 molar mass measurements listed in table 2.

| Sample | $M$/(g/mol) | $u_{rel} \cdot 10^{-9}$ | $n$ |
|---|---|---|---|
| Average S5 | 27.976 970 09(09) | 3.1 | 15 |
| Average S8 | 27.976 970 06(15) | 5.4 | 14 |
| AVO28 boule | 27.976 970 09(15) | 5.4 | 24 |

*2.3. Lattice parameter*

INRIM's combined X-ray/optical interferometer, used to determine the {220} lattice-plane spacing of the enriched silicon crystals, was upgraded and measurements repeated to either confirm the previous result [31] and its uncertainty or to identify possible errors.

First, a 532 nm frequency-doubled Nd:YAG laser was substituted for the 633 nm diode laser. The pressure in the vacuum chamber was also reduced by an order of magnitude, to less than 0.01 Pa. This made any correction for the refractive index of the residual gas essentially inconsequential and ensured a calibration of the optical interferometer with negligible uncertainty.

Next, the delivery, collimation, phase-modulation, and pointing systems of the laser beam were rebuilt to conform to the new wavelength. The beam divergence was reduced, thereby halving the correction for diffraction effects. To have real-time control of the beam pointing, a home-made telescope was placed at the interferometer output port; to ensure stability, it was clamped on the same base plate of the X-ray/optical interferometer.

Then, a plate beam-splitter was substituted for the previously used cube beam-splitter. This ensured that the length difference of the transmitted and reflected light paths was insensitive to any beam translations and rotations. The fixed components of the interferometer – beam splitter, quarter-wave plates, and fixed mirror – were replaced and assembled anew.

In order to produce parallel interfering beams, the interferometer fixed-components were cemented onto a glass plate supported by three piezoelectric actuators. A new power supply, producing sub part-per-million noise and stability, was designed and built to eliminate instabilities between the X-ray and optical interferometers.

PTB also found that the surfaces of the X-ray interferometer crystals were contaminated with Cu, Fe, Zn, Pb, and Ca, caused by the wet etch used to remove any residual surface stress after crystal machining. These contaminants were removed by cleaning the crystals in aqueous solutions of HF and $(NH_4)_2S_2O_8$.

The final upgrade focussed on more accurate temperature measurements. The fixed point cells of INRIM and PTB were compared to establish their temperature difference and to link the INRIM and PTB extrapolations of the lattice parameter and sphere volume to 20 °C. It was not yet possible to verify if the thermometer readings at 20 °C were identical to within the same uncertainty of the fixed-point cell temperatures; this non-uniqueness error was cautiously set to 0.1 mK [32].

To make a reassessment of the measured value and its uncertainty, all the systematic effects were scrutinized and reevaluated with a view to reducing the overall uncertainty and to confirm that the intended goals could be met. Seven surveys of the lattice spacing (with the interferometer crystals aligned as they were originally in the boule and in a reversed arrangement) were carried out from February to June 2014. These surveys were made over 0.95 mm crystal segments centered in 48 different positions. At each position, X-ray counts were recorded in eight different pixels of the interference pattern (11.2 mm high) and then processed using linear regression to obtain 48 values along a line that was the continuation of the laser beam and, therefore, unaffected by Abbe errors. The final average is

$$d_{220}(\text{XINT}) = a(\text{XINT})/\sqrt{8} = 192.014\ 711\ 98(34)\ \text{pm} \qquad (2)$$

where XINT is the X-ray interferometer. At $t_{\text{ITS-90}} = 20$ °C and $p = 0$ Pa, equation (2) expresses the mean lattice spacing value along a strip 46 mm long, orthogonal to the crystal axis, and at a distance of 306 mm from the seed. Details about the measurements and the data analysis together with a discussion of the full error budget are in [33].

As regards the crystal perfection, the NMIJ carried out topographic measurements of the lattice strain in several samples from the AVO28 crystal by means of a self-referenced lattice comparator at the Photon Factory of the High Energy Accelerator Research Organization (KEK, Japan) [34, 35]. The analyser crystal of the X-ray interferometer and a sample, identified by the 4.R1 code and cut from the seed end of the crystal, shows a smooth and homogeneous distribution of lattice spacing values. The standard deviation of the observed variations is $4.9 \times 10^{-9}$ for the 4.R1 crystal. This value is consistent with what is observed by X-ray interferometry, $1.5 \times 10^{-9}\ d_{220}$, with a strain smoothing over $(2 \times 4)$ mm$^2$ areas. By way of contrast, a tail-end sample, identified by the 9.R1 code, shows a two dimensional swirl-like pattern and a greater variability in its lattice spacing values. This observation is consistent with the segregation of impurities into the tail of a crystal purified by the float-zone process. Therefore, the tail sample can be expected to be more contaminated and to display significant variations in its lattice spacing [35].

The mean lattice parameter of each sphere,

$$a(\text{S}) = (1 + \Sigma_i \beta_i \Delta N_i)\ a(\text{XINT}), \qquad (3)$$

was calculated by taking account of the different point-defect concentrations in the spheres and the interferometer. In equation (3), S is sphere AVO28-S5 or AVO28-S8, and XINT is the X-ray interferometer. The subscript $i$ refers to point defects, where $\beta_i$ is the strain coefficient [12, 36] and $\Delta N_i$ is the concentration difference of the point-defect $i$ between the sphere and the interferometer. Unlike the previous determination, where only carbon, oxygen, and boron contamination were considered, the newly measured gradient of nitrogen concentration was also taken into account.

*2.4. Surface*

The surface layer of the Si sphere must be accurately characterized and measured to refine the correction values required for the mass and volume determinations of the spheres. The basic requirements and details of the methods used for surface characterisation are described in [3]. The next section will outline the characterisation techniques, highlighting any changes to the procedures described in [3].

To completely analyze the surface of a Si sphere, a rapid measurement method is required. An ideal technique for investigating a SiO₂ layer on a Si substrate is SE which combines fast measurements (a thickness measurement at a single point in less than 10 seconds) and high precision (repeatability around 10 pm). Using SE, the automatic measurement of a spherical surface with 2600 data points can be completed in 12 hours. Unfortunately SE has two shortcomings: its accuracy (approximately 1 nm) is insufficient for the present application and, as an inverse method, a model of the surface layers must be generated, which then becomes part of the data refinement process. To overcome these limitations, a calibration of the ellipsometric measurement process must be carried out. Surface characterisation thus becomes a two-step process. First, the ellipsometer is calibrated based on reference methods such as XRR at NMIJ and the combination of XRR and XRF analysis at PTB. The second step, the mapping of the surface using ellipsometry, can then proceed.

Subsequent to the last determination of $N_A$, the $^{28}$Si spheres, AVO28-S5 and AVO28-S8, were repolished. The surface layer was therefore modified and, more importantly, simplified when compared to the measurements made in 2009 and 2010. The surficial metallic contaminants (Cu, Ni and Zn) were removed and were therefore no longer part of the surface layer model. For the new $N_A$ determination, the surface layer model illustrated in figure 1 is now applicable.

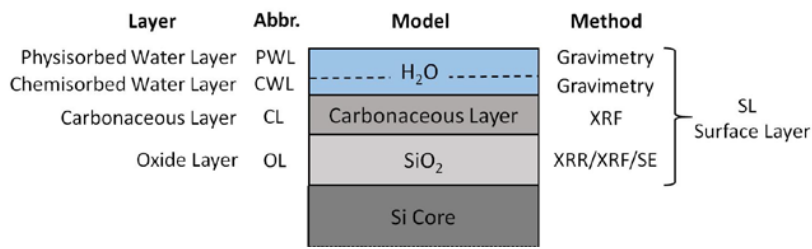

**Figure 1** The surface layers model used for the characterisation of the AVO28 spheres.

The measurements for the surface layer characterisations were performed at NMIJ and PTB. NMIJ has recently installed a new spectral ellipsometer, which enables the automatic measurement of an entire sphere surface, accumulating a large number of data points. PTB used the identical equipment used for the 2010 Avogadro constant determination, which is also capable of automatic surface coverage producing several thousand data points

NMIJ and PTB used different approaches for the calibration of their spectral ellipsometers. The general methodology has already been described in detail in [3]. At the synchrotron radiation laboratory of PTB (BESSY II) [37], direct measurements of the oxide layer (OL) and carbonaceous layer (CL) thicknesses were made using XRF. The calibration of the XRF thickness measurement was done using the ratio of the O-K peak to the Si-L peak of SiO₂ reference samples, whose thickness had been determined by XRR in the vicinity of the oxygen absorption edge at a photon energy of 529 eV. Because these measurements were made in a vacuum, only the chemisorbed water layer (CWL) was present. For the OL thickness determination, it was assumed that all measured O atoms were in the SiO₂ layer. To correct for the CWL layer, the measured thicknesses were adjusted using the mass deposition of the chemisorbed water given in [38] to derive the thickness of the OL layer.

For the determination of the CL layer, the ratio of the C-K peak to the Si-L peak was used and compared to a reference carbon layer whose thickness was again determined by XRR. To derive a thickness for the CL, a mass density of 1.1 g/cm³ was assumed, this being about half of the carbon bulk density. This thickness was required for the NMIJ ellipsometric measurements as input data for the surface layer model used in their data refinement. This thickness was also used in the volume determinations. Since the chemical structure of the CL is highly unpredictable and unclarified, a generous uncertainty was applied to this measurement.

In addition to XRF, XRR was performed directly on the spheres in the vicinity of the O-K edge as described in [3, 39]. The thicknesses determined for the involved layers were in good agreement with the values from XRF. However, as the total layer thickness was well below 2 nm, no oscillations with multiple periods could be observed and the thickness determination was therefore based on previously obtained optical constants, which lead to increased uncertainties.

For the calibration of the ellipsometric measurements at PTB, the OL thickness of a few well defined points on the sphere were determined using the XRR-based XRF measurements. These points could be located using three different markings (cross, "T" and triangle) to an accuracy better than 0.5°. The calibration points were then used for the calibration of the PTB spectral ellipsometer during the mapping of the sphere surface (in effect, an "in-vivo" calibration). By adjusting the alignment of the sphere, a calibration point could be included in each measurement of a great circle. This meant that a short term stability of the instrument of less than 30 minutes was required, which could be expressed as an uncertainty contribution of less than 20 pm. With this "in-vivo" calibration, requirements for the simulation model for the refinement of the ellipsometric data were dramatically simplified because the influence of the CWL and CL layers are inherently included in the calibration constant $C$. Only the linearity of the ellipsometer and the homogeneity of the CWL and the CL layers were required for the simulation model.

The calibration measurements of the spheres at the PTB synchrotron radiation laboratory were carried out in November 2013 and January 2014 (AVO28-S5c) and in January 2014 and July 2014 (AVO28-S8c), respectively. The ellipsometric mapping of sphere AVO28-S5c was done with 5184 data points, while sphere AVO28-S8c had 15 552 data points. The results of these measurements are listed in tables 4a and 4b.

NMIJ used a different approach for the calibration of their ellipsometer. The oxide layer thickness on the two spheres was measured by a spectroscopic ellipsometer at NMIJ. The reliability of the spectroscopic ellipsometer was checked by using $SiO_2$ layers on Si wafers with thicknesses certified by XRR [3]. In the previous paper [3], the oxide layer thickness was measured at only 20 points on the surface of the spheres. To increase the number of measurement points for improving the reliability of the measurement, a new spectroscopic ellipsometer equipped with an automatic sphere rotation system was developed [40]. This new instrument was a spectral ellipsometer with a rotating polarizer. Its spectral bandwidth ranged from 250 nm to 990 nm. The Si sphere was placed on two rollers made of PEEK (Polyether ether ketone) and could be rotated by the automatic sphere rotation system around both its vertical and horizontal axes. The sphere rotation system was integrated into the spectroscopic ellipsometer, thereby enabling the automatic mapping of the oxide layer thickness over the entire surface of the sphere. The measurements were based on 2594 points, regularly distributed over the entire sphere surface. The sphere surface was subdivided into small cells of equal area and the measurement points were distributed uniformly to each cell.

The oxide layer thickness of AVO28-S8c was measured in June 2014 and that of AVO28-S5c was measured in July and September 2014. Although the 2594 points were distributed almost uniformly over the sphere surface, the uniformity was not perfect. To estimate the effect of this non-uniform distribution of the measurement points on the average value of the oxide layer thickness, the measurements at the 2594 points were repeated 3 times. Between each set of measurements, the sphere was rotated to distribute 7782 (=2594×3) points as uniformly as possible. Before each set of measurements, the spheres were washed using the same procedure used for the international comparisons of the mass and diameter of a Si sphere in the International Avogadro Coordination project [38, 41]. The standard deviation of the mean OL thickness for 3 sets of the measurements was less than 0.1 nm, showing the uniformity of the distribution of the 2594 points and the reliability of the measurement system.

The ellipsometric data were analyzed at NMIJ based on the surface model with four layers to evaluate the $SiO_2$ thickness. Since the ellipsometric measurement was performed in air, the model consists of a carbonaceous layer (CL), a physisorbed water layer (PWL), a chemisorbed water layer (CWL) and an oxide layer (OL). The thickness of PWL was estimated to be 0.39(9) nm and 0.43(9) nm for AVO28-S5c and AVO28-S8c, respectively. These results were obtained at NMIJ from the comparison weighings in nitrogen gas of about 1200 Pa and in water vapour of about 1200 Pa [42] for the two spheres using a stainless steel weight as a reference. The amount of PWL on the stainless steel weight was determined in advance by comparison weighings using artifacts with large surface area difference. The thickness of the CWL was estimated from the published value of Mizushima [43] to be 0.28(8) nm. The thickness of the CL was measured by XRF at PTB to be 0.60(18) nm and 0.49(16) nm for AVO28-S5c and AVO28-S8c, respectively, on the assumption that the density of the CL is 1.1 g/cm$^3$.

To evaluate the OL thickness, the measured ellipsometric parameters were fitted by the aforementioned four-layer model, fixing all the sub-layer parameters, except the thickness of the OL. The evaluated OL thicknesses were 0.76 (27) nm and 0.64(25) nm for AVO28-S5c and AVO28-S8c, respectively. The thickness of each layer is summarised in table 4a. The dominant uncertainty source for the OL thickness determination is the thickness of the CL.

**Table 4a.** The thickness of the surface layer and its constituent sub-layer components, $d_{CL}$, $d_{CWL}$, $d_{PWL}$ and $d_{OL}$. See figure 1 for the key to the subscript abbreviations.

| Sphere | Lab. | $d_{CL}$/nm [1] | $d_{CWL}$/nm [2] | $d_{PWL}$/nm [3] | $d_{OL}$/nm | Date of measurement (OL) | $d_{SL}$/nm [4] |
|---|---|---|---|---|---|---|---|
| AVO28-S5c | PTB | 0.60(18) | 0.28(8) | - | 0.91(14) | Jan. 2014 | **1.79(24)** |
| | NMIJ | 0.60(18) | 0.28(8) | 0.39(9) | 0.76(27) | July and Sep. 2014 | **1.64(33)** |
| | average | 0.60(18) | 0.28(8) | | 0.88(12) | | **1.76(23)** |
| AVO28-S8c | PTB | 0.49(16) | 0.28(8) | - | 1.17(13) | July 2014 | **1.94(22)** |
| | NMIJ | 0.49(16) | 0.28(8) | 0.43(9) | 0.64(25) | June 2014 | **1.41(31)** |
| | average | 0.49(16) | 0.28(8) | | 1.06(22)[5] | | **1.83(28)** |

[1] The thickness of the CL measured by XRF at PTB was based on the assumption that the density of the CL was 1.1 g/cm$^3$. The uncertainty of this thickness was estimated using the surface analysis results from the previous measurement [3].
[2] The $d_{CLW}$ was calculated from data reported by Mizushima [43].
[3] The data for the $d_{CWL}$ came from comparison weighings of the two spheres in nitrogen gas, at a pressure of ca.1200 Pa, and in water vapour, at a pressure of ca. 1200 Pa [42]. The density of the PWL was assumed to be 1.0 g/cm$^3$.
[4] This value does not include the thickness of the PWL.
[5] The Birge ratio of the thickness values of the oxide layer of AVO28-S8c is 1.8. Therefore, the uncertainty of the weighted mean was multiplied by 1.8. A possible reason for the difference in the oxide layer determinations may be that NMIJ used the CL thickness value of PTB and the surface cleaning status were not identical at PTB and NMIJ.

**Table 4b.** Mass of the surface layer and its constituent sub-layers. See figure 1 for the key to the subscript abbreviations.

| Sphere | Lab | $m_{CL}$/µg | $m_{CWL}$/µg | $d_{PWL}$/nm | $m_{OL}$/µg | Date of measurement (OL) | $m_{SL}$/µg * |
|---|---|---|---|---|---|---|---|
| AVO28-S5c | PTB | 16.6(5.7) | 7.7(2.2) | - | 55.2(8.9) | Jan. 2014 | **79.5(10.9)** |
| | NMIJ | 16.6(5.7) | 7.7(2.2) | 10.8(2.5) | 46.1(16.5) | July and Sep. 2014 | **70.4(17.7)** |
| | average | 16.6(5.7) | 7.7(2.2) | | 53.4(7.7) | | **77.7(10.0)** |
| AVO28-S8c | PTB | 13.5(5.2) | 7.7(2.2) | - | 71.0(8.5) | July 2014 | **92.2(10.2)** |
| | NMIJ | 13.5(5.2) | 7.7(2.2) | 11.9(2.7) | 38.9(15.3) | June 2014 | **60.0(16.3)** |
| | average | 13.5(5.2) | 7.7(2.2) | | 64.3(13.7) | | **85.5(14.8)** |

* The mass of the PWL was not included in this value.

### 2.5. Volume

Optical interferometers were utilised to determine the volume of the two spheres by measuring the diameters of the spheres and calculating their volumes. Although the elementary dimensional measurements were based on the same principle, different types of interferometers with different optical set-ups were used. A precise measurement typically takes advantage of a differential approach. Thus, the measurement of a sphere proceeds in two steps: first is the measurement of the dimensions of a stable optical etalon, $D$. The second step then involves the insertion of a sphere into the etalon and the measurement of the gaps between the sphere and the etalon, $d_1$ and $d_2$. The diameter of the sphere is calculated from the difference of these measurements $d = D - d_1 - d_2$. To fully characterise a sphere, the diameter is measured in many different directions.

NMIJ measured 1450 and 870 diameters of AVO28-S5c and AVO28-S8c, respectively, using an improved optical interferometer with a flat etalon [40]. The $^{28}$Si sphere was placed between the two flat etalon plates, and $d_1$, $d_2$ and $D$ were measured by phase shifting interferometry with optical frequency tuning. The main improvements from the previous work [44] are summarized below.

Firstly, the geometrical shapes of the optical components were optimized. The largest uncertainty source in the previous volume measurement at NMIJ was the analysis of the interference fringes [44]. An analysis using the ray-tracing method showed that a possible cause for the uncertainty in the analysis of the interference fringes is the multiple reflection of the beam between the tilted surface of the etalon and the sphere surface. An increased tilt of the etalon surface was therefore expected to reduce the effect of the multiple reflections. Based on this analysis, a new etalon with a larger wedge angle was installed [40].

Secondly, the uniformity of the diameter measurement directions was improved. One of the major uncertainty sources in the previous volume measurement was the experimental standard deviation of the mean volume. The diameter measurement directions were based on 70 directions and the non-uniform distribution of these 70 directions was estimated to increase the experimental standard deviation of the mean volume [44]. To decrease this uncertainty, a strategy to distribute the measurement directions as uniformly as possible was developed [40]. The new distribution consists of 145 directions. The sphere surface is subdivided into small cells of equal area and the measurement points are distributed uniformly to each cell.

Thirdly, the optical frequency standard was upgraded. An iodine-stabilized He-Ne (HeNe/$I_2$) laser was used as the optical frequency standard in the NMIJ interferometer [44]. However, the HeNe/$I_2$ laser was highly sensitive to acoustic noise. An optical frequency comb was therefore employed as the standard to obtain more precise and reliable volume measurements [40]. This light source system is much more robust than the HeNe/$I_2$ laser and can be operated for a long period of time such as 10 days. The relative uncertainty of the frequency of the laser is estimated to be approximately $1 \times 10^{-11}$ at 1 s averaging time. This uncertainty is limited by the statistical frequency fluctuation of the offset laser and can be ignored for the volume measurement.

The measured diameter is the 'apparent' diameter, which is not corrected for the phase shift due to the surface layer. The mean apparent diameters at 20.000 °C and 0 Pa are 93 710.811 11(62) µm and 93 701.526 29(68) µm for AVO28-S5c and AVO28-S8c, respectively. The relative standard uncertainties of the volume measurement are $2.0 \times 10^{-8}$ and $2.2 \times 10^{-8}$ for AVO28-S5c and AVO28-S8c, respectively. Table 5 shows the uncertainty budget for the determination of the apparent volumes. In the previous work, the volumes of the two spheres were determined by NMIJ with relative standard uncertainties of $5.0 \times 10^{-8}$ and $4.4 \times 10^{-8}$ for AVO28-S5 and AVO28-S8, respectively [44]. By the improvements described above, the uncertainty contributions from the interferogram analysis and the random component were decreased, resulting in the significant reduction in the uncertainty of the volume measurement. The dominant uncertainty source at present is the correction of the diffraction effect on the diameter measurement [45]. The value of this correction was estimated to be 0.45(50) nm.

**Table 5.** The relative uncertainty budget for apparent volume measurements of the $^{28}$Si spheres at NMIJ [44].

|  | $(u(V)/V)/10^{-9}$ | |
| --- | --- | --- |
|  | Avo28-S5c | Avo28-S8c |
| Interferogram analysis | 10.3 | 10.3 |
| Temperature | 4.8 | 4.8 |
| Diffraction effect | 16 | 16 |
| Standard deviation of the mean volume | 3.3 | 9.6 |
| **Total** | **20** | **22** |

The PTB interferometer was based optically on a completely spherical geometry [46]. This means that the reference faces were spherical, so that the etalon forms opposing segments (caps) which surround the sphere. Furthermore, the illuminating light wave was converted by a set of objectives into a focused beam, so that these conical rays would hit the reference face and the sphere perpendicularly. Thus, the relationship $d = D - d_1 - d_2$ would be true for all points $\theta, \varphi$ within the field of view (covering 60°). This enabled the acquisition of high resolution topographies of the sphere. Each sphere was provided with three different marks (following the orientation of the crystallographic axes) so that a sphere could be initially oriented by one mark with the help of the interferometer camera. Subsequently it was oriented by means of the high-resolution encoder equipment of the sphere positioning motors. Each measurement, at all orientations of the sphere, could then be related to its absolute position on the sphere [47].

The volume of this 'sphere' (see figure 2) was calculated by considering the areal weighting of each measuring point [48]. To eliminate any possible stability problems of the etalon and to monitor the stability of the interferometer, PTB always alternated measurements between sphere and empty etalon.

The dimensional measurement is traced back to the laser wavelength of a 633 nm He-Ne laser recommended by BIPM. The stability of the stabilised He-Ne laser, the auxiliary unmodulated He-Ne laser and the extended-cavity-diode-laser (ECDL) for the wavelength-tuning was at the $10^{-12}$ level and therefore did not contribute to the volume uncertainty budget.

For the measurements used in this paper, the interferometer was improved by stabilizing the irradiance of the laser light. The light from the ECDL was split into two parts and guided to the two arms of the interferometer by multi-mode fibres. Two aspects of these measurements were taken into account: a mode scrambler provided a uniform intensity distribution within the field of view, and a monitor photo-diode combined with a 'noise eater' (i.e. a servo control with a fast liquid crystal light modulator) stabilised the intensity of the interferometer input [49].

Sphere AVO28-S5c was measured in August 2013. It was washed, following the suggestions of the IAC, with distilled water and deconex OP162, a pH neutral, salt-free cleaning concentrate for precision optical components. After extensive purging with distilled water, the sphere was rinsed with alcohol (p.A., pro analysi, analytically pure). The measurement followed a 50%-overlapping procedure: with a field of view of 60°, the sphere was rotated for the next measurement by only 30°. The sphere was measured from both sides, so that a complete set of diameters measured through arm 1 and a complete set measured through arm 2 of the interferometer were derived. For each single measurement, the temperature was corrected by a recently calibrated Pt25 resistance thermometer as well as with a set of thermocouples [50].

Sphere AVO28-S8c was measured in a like manner in September 2014. In this case, the data came from two sets of completely overlapping measurements, each taken in a different sequence and with different rotation steps. Nevertheless, the volume characterization was based on $7 \times 10^5$ diameter values.

Compared to the initial spherical state produced by the Australian polishing, the PTB polishing removed about 12 µm from the diameter of sphere AVO28-S5. The p-v value of the diameter was reduced from 98 nm to 69 nm, illustrating nicely the effects of anisotropy of the modulus of elasticity (E-modulus) of a silicon single crystal [51, 52]. For AVO28-S8, two previous measurement cycles, AVO28-S8a and AVO28-S8b, have already been reported [5, 12, 48]. The diameter of the sphere AVO28-S8 has now been reduced by 20 µm and its p-v value decreased from 90 nm to 38 nm.

The uncertainties follow the considerations and calculations presented in [48]. Due to the smaller deviations from roundness (decreased p-v values) for the repolished spheres, the influence of the wave front aberrations were also presumably reduced (table 6), but such distortions remain the principal uncertainty contribution for the present sphere interferometer at PTB. A new interferometer, with a set of objectives with considerably reduced wave front aberrations, is currently being tested. Results for this new instrument together with optical simulation calculations will be reported soon.

**Table 6.** Relative uncertainty budgets for the apparent volume measurements of the $^{28}$Si spheres at PTB [48].

|  | $(u(V)/V)/10^{-9}$ | |
|---|---|---|
|  | AVO28-S5c | AVO28-S8c |
| Interferometry | 4 | 4 |
| Temperature | 6 | 6 |
| Wavefront distortions | 25 | 19 |
| Parasitic interferences | 0.2 | 0.2 |
| Volume | 6 | 6 |
| **Total** | **27** | **21** |

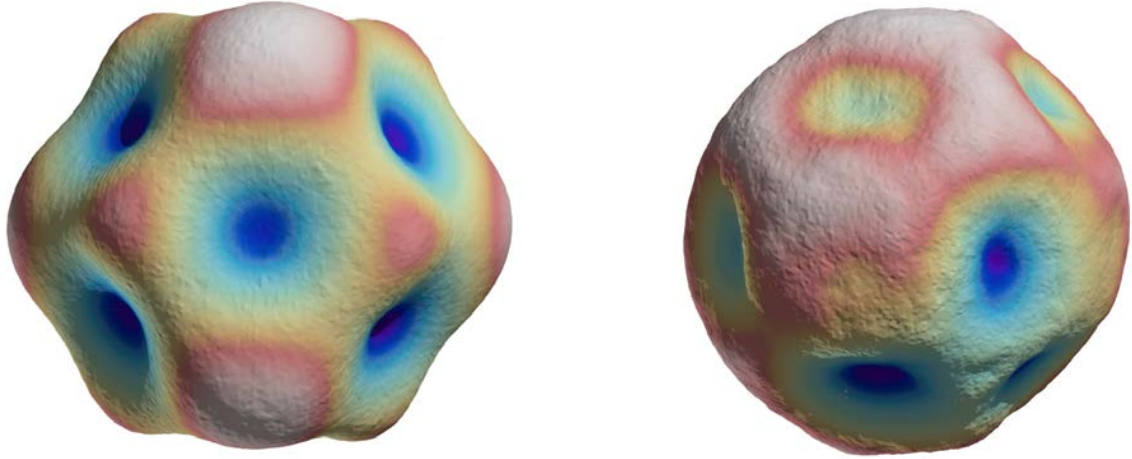

**Figure 2.** Diameter topographies of the $^{28}$Si spheres AVO28-S5c (left, (p-v)$_{diameter}$ = 69 nm) and AVO28-S8c (right, (p-v)$_{diameter}$ = 38 nm).

To take into account the optical behaviour of the surface layers, a layer model was created [48] and is illustrated in figure 1. The optical constants of the different surface layers are given in table 7.

**Table 7.** The optical constants (refractive index $n$ and the absorption index $k$) of the surface layers (SiO$_2$, H$_2$O and C$_m$H$_n$) and the silicon core (Si). The Abbr column refers to the abbreviations of the model layers in figure 1.

| Layer | Abbr | $n$ | $k$ |
|---|---|---|---|
| Si | Si Core | 3.881(1) | 0.019(1) |
| SiO$_2$ | OL | 1.457(10) | 0 |
| H$_2$O | CWL, PWL | 1.332(10) | $1.54(1.00) \times 10^{-8}$ |
| C$_m$H$_n$ | CL | 1.45(10) | 0(0.1) |

Table 8 lists the diameter values and the phase shift corrections $\Delta\phi$ due to the optically applicable layers for both participating institutes (different types of interferometers). Calculations using the layer model suggested that the influence of the surface layers on the optical diameter measurements was quite small. The value of the phase shift showed a zero-crossing for thicknesses in the range of 1.5 nm to 2 nm.

**Table 8.** Diameter and volume of $^{28}$Si spheres ($t_{\text{ITS-90}}$ = 20 °C, vacuum).

| Sphere | Lab. | Mean apparent diameter/nm | Date of measurement | $\Delta\Phi$/nm[1] | Mean diameter of Si core/nm | Volume $V$ of Si core/cm$^3$ |
|---|---|---|---|---|---|---|
| AVO28-S5c | PTB | 93 710 811.38(83) | August 2013 | 0.000(17) | 93 710 811.38(83) | 430.891 291(12) |
| AVO28-S5c | NMIJ | 93 710 811.11(62) | June - July 2014 | -0.001(25) | 93 710 811.11(62) | 430.891 288(9) |
| AVO28-S5c | weighted mean | 93 710 811.21(50) | | | 93 710 811.21(50) | 430.891 289(7) |
| AVO28-S8c | PTB | 93 701 526.24(66) | September 2014 | 0.010(32) | 93 701 526.26(66) | 430.763 222(9) |
| AVO28-S8c | NMIJ | 93 701 526.29(68) | Nov. - Dec. 2014 | -0.009(28) | 93 701 526.27(68) | 430.763 223(9) |
| AVO28-S8c | weighted mean | 93 701 526.26(47) | | | 93 701 526.26(47) | 430.763 223(7) |

[1] $\Delta\phi$ = phase shift corrections

*2.6. Mass*

The masses of the spheres AVO28-S5c and AVO28-S8c were determined in air and under vacuum by the BIPM, NMIJ and PTB.

All measurements at the BIPM were carried out using their Sartorius CCL 1007 mass comparator. Before weighing, the two spheres were cleaned three times, using the cleaning procedure recommended by the National Metrology Institute of Australia, NMIA [53]. A set of stainless steel air buoyancy artefacts, consisting of a tube (Cp) and a hollow cylinder (Cc), was used to determine the air density for the buoyancy correction. A set of Pt-Ir sorption artefacts, consisting of a cylinder (A0) and a stack of 8 disks (A18), was used to establish the link between masses in air and vacuum. The masses in air were measured three times, with weighings in vacuum occurring in-between. After each change of condition, a stabilization period of at least 3 days was observed before starting the next set of measurements under the new conditions.

The weighings of the two spheres were carried out directly after the first phase of the Extraordinary Calibrations against the IPK. The spheres were compared in air with BIPM working standard No. 77, which had itself been weighed against working standards Nos. 91 and 650. The two latter working standards had already been compared directly with the IPK, in air. Tables 9 and 10 give the masses of both spheres, in air and in vacuum, with respect to the mass of the IPK. During the Extraordinary Calibrations, it was observed that the BIPM 'as-maintained' mass unit was 35 µg smaller than the mass of the IPK. The standard uncertainty of this mass difference is estimated as 3 µg. Therefore, the masses of AVO28-S5c and AVO28-S8c, expressed in the mass unit maintained formerly at the BIPM, are 35 µg larger than the values given in tables 9 and 10.

A study was undertaken at the BIPM to determine the mass of the CWL present on the surface of a natural silicon sphere. Two methods were used: baking the sphere under vacuum and immersing the sphere in doubly distilled water [54]. For both methods, the chemical adsorption coefficient was obtained by determining the mass difference under vacuum conditions (to reduce the uncertainty) prior to and after placing the sphere in air, in order to reintroduce the chemisorbed water to the surface of the sphere. The mean chemical adsorption coefficient thus obtained was 0.026 µg cm$^{-2}$ with a standard uncertainty ($k = 1$) of 0.012 µg cm$^{-2}$. The BIPM results confirmed those obtained by NMIJ/AIST (Japan) which had measured the adsorption isotherms on $SiO_2$/Si(100) plane surfaces [43].

NMIJ conducted mass measurements using an early version of the Mettler-Toledo M_one mass comparator [55]. The measurements under vacuum showed a higher reproducibility than those in air, where air buoyancy and convection effects affected the weighing stability. The standard uncertainty of the mass difference measurement under vacuum was about 5 µg.

The silicon spheres were washed manually by rubbing their surfaces with nitride rubber gloves using a neutral detergent for optical components, followed by a rinse with 30 dm$^3$ of pure water and 1 dm$^3$ of ethanol. The washing procedure is basically the same as that described in the "protocol for the international mass comparison on the two $^{28}$Si spheres" distributed to the participants in April 2008 [53]. The protocol does not clearly specify the number of washings. Therefore, NMIJ investigated the washing effects on reproducibility by washing AVO28-S5c three times and by washing AVO28-S8c twice. The repeated washings, with subsequent mass measurements at NMIJ, showed a mass decrease between 3 µg and 8 µg after each washing. This phenomenon could be interpreted as a decrease of a carbonaceous layer (CL) on surfaces by the washing procedure. The result of the washing experiment suggests the number of washings specified in the protocol should be more than three to achieve a stable mass value in a few microgram level. It should be also noted that the washing effect would be compensated, to a great extent, by the surface characterisation described in section 2.4 if we specify the washing procedure properly. This time, NMIJ estimated the standard uncertainty of the reproducibility of the washing procedure specified in the aforementioned protocol to be 4 µg, assuming uniform distribution with a half width of 7 µg.

The traceability of NMIJ's mass value to the IPK was achieved by using the mass value 1 kg + 0.360 mg for the Pt-Ir prototype No. 94 (BIPM Certificate No. 59, 2009) measured at the BIPM in August 2009 and also by applying a correction of –0.0301 mg, which corresponded to the value during this time period recommended by the BIPM in December 2014 [8]. In addition, the mass value 1 kg + 0.176 mg for the Pt-Ir prototype No. 6 (BIPM Certificate No. 8, 1993) at the third periodic verification of national prototypes of the kilogram was used to determine the long-term drift rate of NMIJ's mass value. The results for the weighings are shown in tables 9 and 10. The NMIJ measurement results are the average obtained by two measurement cycles.

**Table 9.** Masses of the $^{28}$Si spheres as measured in air (without a correction for the reversible water layer).

| Sphere | Laboratory | Mass/kg | Mass unc./µg | Date of measurement |
|---|---|---|---|---|
| AVO28-S5c | PTB | 0.999 698 447 | 12 | Nov./Dec. 2013 |
| | BIPM | 0.999 698 423 | 13 | Feb./March 2014 |
| | NMIJ | 0.999 698 437 | 16 | May/June 2014 |
| AVO28-S8c | BIPM | 0.999 401 303 | 13 | Feb./March 2014 |
| | NMIJ | 0.999 401 328 | 13 | Apr./May 2014 |
| | PTB | 0.999 401 325 | 12 | Oct./Nov. 2014 |

**Table 10.** Masses of the $^{28}$Si spheres as measured in vacuum.

| Sphere | Laboratory | Mass/kg | Mass unc./µg | Date of measurement |
|---|---|---|---|---|
| AVO28-S5c | PTB | 0.999 698 438 5 | 6.9 | Nov./Dec. 2013 |
| | BIPM | 0.999 698 430 1 | 4.4 | Feb./March 2014 |
| | NMIJ | 0.999 698 437 3 | 7.6 | May/June 2014 |
| | Weighted mean | 0.999 698 433 2 | 3.5 | |
| AVO28-S8c | BIPM | 0.999 401 309 5 | 4.3 | Feb./March 2014 |
| | NMIJ | 0.999 401 320 9 | 7.8 | Apr./May 2014 |
| | PTB | 0.999 401 316 0 | 7.3 | Oct./Nov. 2014 |
| | Weighted mean | 0.999 401 312 8 | 3.5 | |

At PTB, the mass determinations of the $^{28}$Si spheres AVO28-S5c and AVO28-S8c were performed in air and under vacuum during the periods November/December 2013 and October/November 2014, respectively. The procedure agreed to in [53] for the Avogadro project was used for cleaning the spheres. After cleaning, the measurements were made on a Mettler-Toledo M_one mass comparator using the prototype of the kilogram No. 70 and the Pt-Ir mass standard PtSk-Z as reference masses. Table 11 shows the traceability path between the reference masses used for the mass determination of the $^{28}$Si spheres and the prototypes of the kilogram at the BIPM. The last calibration of prototype No. 70 at the BIPM was performed in June 2013 (Certificate No. 70, BIPM 2013). The BIPM determined the mass of this prototype to be 1 kg – 0.207(7) mg. Mass comparisons with other prototypes of the kilogram were made at PTB before and after the prototype was hand-carried between PTB and BIPM. The results showed a mass loss of 5 µg. Because it is unknown whether this mass change occurred before or after the calibration of the prototype, this mass difference had to be considered in the uncertainty budget as an uncertainty contribution due to the instability of the mass of the prototype.

**Table 11**. The traceability path between the reference masses used for the mass determination of the $^{28}$Si spheres at PTB and the prototypes of the kilogram at the BIPM

| Sphere | Date | Reference masses | | | | | |
|---|---|---|---|---|---|---|---|
| | | Name | Last calibration by PTB | Name | Last calibration by PTB | Name | Last calibration by BIPM |
| AVO28-S5c | Nov./Dec. 2013 | PtSk-Z | Nov. 2013 (against No. 70) | | | No. 70 | Jun. 2013 |
| AVO28-S8c | Oct./Nov. 2014 | PtSk-Z | Oct. 2014 (against No. 70) | | | | |
| | | No. 70 | Oct. 2014 (against No. 52) | No. 52 | Nov. 2013 (against No. 70) | No. 70 | Jun. 2013 |

The Pt-Ir cylinder PtSk-Z is one of two Pt-Ir artefacts used as sorption artefacts for the determination of the sorption correction and as a link between the mass of the silicon spheres in vacuum and the prototype of the kilogram No. 70 in air [53]. The surface area difference between the artefacts amounts to 183 cm². In order to apply a buoyancy correction in air with the lowest uncertainty, the air density was measured using buoyancy artefacts [53].

For the sphere AVO28-S5c, a mass of 0.999 698 483(13) kg and 0.999 698 4743(93) kg was determined in air (without correction of reversible sorption) and in vacuum, respectively. For sphere AVO28-S8c, a mass of 0.999 401 362(14) kg was determined in air (again without correction of reversible sorption) and 0.999 401 3534(94) kg in vacuum. For both spheres, the mass difference between the measurements in air and in vacuum amounted to about 8.5 µg. This value corresponds to a change of the sorption coefficient between vacuum and air of 30 ng/cm², which is in good agreement with the values published in [56] and [57].

In December 2014, PTB was informed by the BIPM about revised mass values for BIPM calibrations following the Extraordinary Calibration Campaign against the IPK [8]. Consequently, revised mass values and revised drift corrections of the involved prototypes, No. 70 and No. 52, had to be considered. Regarding the calibration of prototype No. 70 in June 2013, a corrected mass value of 1 kg - 0.242(3) mg was assigned by the BIPM. The difference between the original and the corrected mass value amounts to 35 µg. In due consideration of the traceability path given in table 11, the mass values determined for the $^{28}$Si spheres AVO28-S5c and AVO28-S8c at PTB were revised correspondingly (tables 9 and 10).

The weighted mean of the masses measured in vacuum was used for the determination of the Avogadro constant (table 10). The effect of the correlation arising from the common traceability of all masses to the IPK was taken into account for the weighted mean and its uncertainty, although its magnitude is nearly negligible.

To determine the mass of the silicon core, the mass of the surface layers was subtracted from the mass of the sphere. In addition, owing to point defects, there is a difference between the mass of a sphere having Si atoms occupying all regular sites and the measured mass value.

$$m_{\text{deficit}} = V \Sigma_i (m_{28} - m_i) N_i \quad (4)$$

In equation (4), $m_{28}$ and $m_i$ are the masses of a $^{28}$Si atom and of the point defect named $i$, respectively (a vacancy mass is zero.) Oxygen was associated with an interstitial lattice site, so that $m_O$ is the sum of the oxygen and $^{28}$Si masses. The same applies to nitrogen impurities. $V$ is the sphere volume and $N_i$ is the concentration of the point defect $i$ (see section 2.1).

The etching did not completely remove the metals from the AVO28-S5 sphere. Approximately 5% of the original metallic contaminant remained (corresponding to 4 µg) and had probably diffused into the sphere during the thermal oxidation that was performed before repolishing. After repolishing the spheres, no metals could be detected by XRF at the sphere's surface. Therefore, a mass correction of -4(3) µg was added to the mass deficit. A mass deficit of 3.8(3.8) µg was then calculated for the AVO28-S5c sphere and a mass deficit of 22.7(3.5) µg for the AVO28-S8c sphere.

After the surface-layer mass ($m_{\text{SL}}$, table 4b in section 2.4) was subtracted and the mass deficit ($m_{\text{deficit}}$) was added, the Si core mass $m = m_{\text{sphere}} - m_{\text{SL}} + m_{\text{deficit}}$ could be calculated for both spheres (table 12).

## 3. Avogadro constant

Table 12 gives the results of the new and refined measurements of the molar mass, lattice parameter, volume, mass, and density of the cores of the enriched silicon spheres. To evaluate and express the measurement uncertainties, the approach recommended by the Guide to the Expression of Uncertainty in Measurement [58] was applied using the GUMWorkbench software [59]; covariances were calculated and folded into the uncertainty analysis. The quantities dominating the total uncertainty of the revised Avogadro constant $N_A$ were the apparent diameter, $D_m$, of the spheres and the mass of the surface layer, $m_{\text{SL}}$ (see table 13).

**Table 12.** $N_A$ determination. The lattice parameter, volume, and density were measured at $t_{ITS-90} = 20.0$ °C and $p = 0$ Pa.

| Quantity | Unit | AVO28-S5c | AVO28-S8c |
|---|---|---|---|
| $M$ | g/mol | 27.976 970 09(15) | 27.976 970 09(15) |
| $a$ | pm | 543.099 6219(10) | 543.099 6168(11) |
| $V$ | cm$^3$ | 430.891 2891(69) | 430.763 2225(65) |
| $m$ | g | 999.698 359(11) | 999.401 250(16) |
| $\rho = m/V$ | kg/m$^3$ | 2320.070 943(46) | 2320.070 976(51) |
| $N_A$ | $10^{23}$ mol$^{-1}$ | 6.022 140 72(13) | 6.022 140 80(14) |

The new $N_A$ determinations, based on a careful reanalysis of the two AVO28 spheres, are summarized in table 14; they differ by only $13(20) \times 10^{-9} N_A$. Averaging these two values, the final value for the Avogadro constant becomes

$$N_A = 6.022\ 140\ 76(12) \times 10^{23}\ \text{mol}^{-1}, \tag{5}$$

with a relative standard uncertainty of $20 \times 10^{-9}$.

**Table 13.** The uncertainty budget for the new $N_A$ determination using AVO28-S5c. The percent contributions to the total uncertainty are the relevant variance fractions ratioed to the total variance. The principal uncertainty contributions are, at present, due to surface characterization and the volume determination.

| Quantity | Relative uncertainty/$10^{-9}$ | Contribution/% |
|---|---|---|
| Molar mass ($M$) | 5 | 6 |
| Lattice parameter ($a$) | 5 | 6 |
| Surface characterization ($m_{SL}$) | 10 | 23 |
| Sphere volume ($V$) | 16 | 59 |
| Sphere mass ($m$) | 4 | 4 |
| Point defects | 3 | 2 |
| Total | 21 | 100 |

**Table 14.** Value of the Avogadro constant based upon the repolished $^{28}$Si-enriched silicon spheres.

| Sphere | $N_A/10^{23}$ mol$^{-1}$ | $u_r/10^{-9}$ |
|---|---|---|
| AVO28-S5c | 6.022 140 72(13) | 21 |
| AVO28-S8c | 6.022 140 80(14) | 23 |
| Mean value | 6.022 140 76(12) | 20 |

## 4. Conclusions

The value $N_A = 6.022\,140\,82(18) \times 10^{23}$ mol$^{-1}$ given in a previous paper [12] must also be updated due to the re-calibration of the mass standards following the recent Extraordinary Calibration Campaign against the IPK. A provisional corrected value would be $N_A = 6.022\,140\,99(18) \times 10^{23}$ mol$^{-1}$.

Some considerations of the potential correlations between the different measurement components contributing to the new and previously reported $N_A$ values should be addressed. The molar mass measurements are uncorrelated; they were amply repeated in three different laboratories using different calibration and measurement approaches as well as TMAH instead of NaOH as matrix diluent. The measurements of the lattice parameters differed in significant respects: the optical interferometer was rebuilt to accommodate a different wavelength, the temperature measurements relied on new equipment and new calibrations, and the apparatus was completely disassembled and realigned. However, because the same interferometer crystal was used for both measurements, and the lattice parameter values relied on measurements from a single laboratory, the past and present results may be correlated by up to 50%. Because the AVO28-S5 and AVO28-S8 spheres were fully repolished, the contributions of their surface properties to the error budget are not correlated. The same applies to the volume measurements; however, diffraction effects – which are extremely difficult to model or investigate – may be correlated between the old and new values. Mass measurements were newly calibrated and traceable back to the IPK. Therefore, their correlation with the past measurements is considered to be negligible. Finally, the point defect contributions to the error budget are basically from the same source and are thus 100% correlated. Details will be published in a separate paper [60].

The forthcoming definition of the kilogram [61] will be based on fixing the value of the Planck constant. To compare this new $N_A$ value with the most recent Planck constant results by the watt balance (WB) experiments [62, 63], we can convert them to an equivalent Avogadro constant value using the molar Planck constant ($N_A h = 3.990\,312\,7176(28) \times 10^{-10}$ J s mol$^{-1}$), which has a relative standard uncertainty of only $0.7 \times 10^{-9}$ [64]. The NRC WB value was updated for the recalibration of mass standards following the Extraordinary Calibration Campaign against the IPK, $h(\text{NRC}) = 6.626\,070\,11(12) \times 10^{-34}$ Js [65]. The NIST updated their former $h$ values and published a combined $h$ value for the NIST-3 WB, $h(\text{NIST-3}) = 6.626\,069\,36(37) \times 10^{-34}$ Js [66]. Note however that we have been unable to update the CODATA 2010 $N_A$ value. These results are compared in figure 3. The accuracy and uncertainty of the new determination of $N_A$ (equation (5)) is within the targeted relative uncertainty so as to make the kilogram redefinition possible [61]; therefore, this new measurement result demonstrates a successful *mise en pratique* of a definition based upon a fixed value of the Planck constant.

Because clear-cut effects of crystal imperfections have not yet been detected, the uncertainty of equation (5) still appears to be limited by the performance of the measurement instrumentation. Therefore, provided the source material is chemically and physically well characterized with respect to the mass fractions of the minor isotopes and impurities, the value of the lattice parameter, and the crystallographic perfection, material realizations of the kilogram and its submultiples in the form of crystal artefacts require only volume measurements and surface characterizations. These same two parameters would also be the only two quantities necessary to monitor the secular stability of the artefacts.

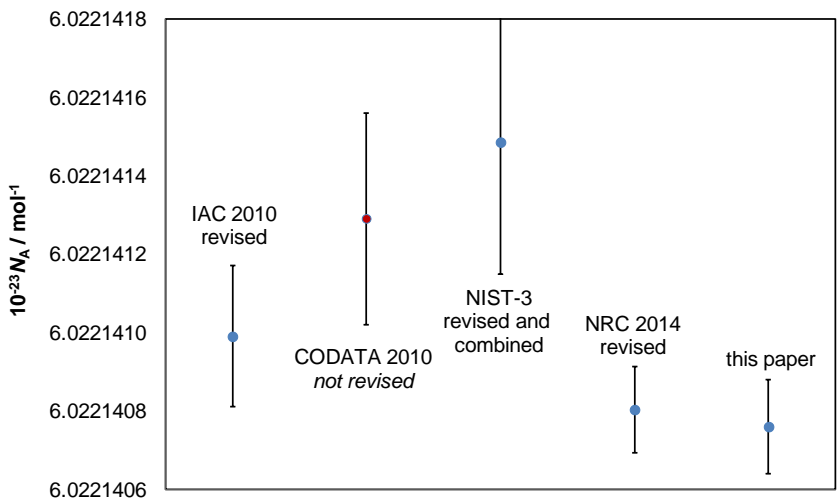

**Figure 3.** The most accurate Avogadro constant determinations available at present. The value labelled CODATA 2010 is not corrected for the recalibration of the mass standards following the Extraordinary Calibration Campaign against the IPK. The vertical error bars indicate the standard uncertainty of each value.

**Acknowledgements**

The authors would like to thank all colleagues involved in this project, in particular Alain Picard, Michael Hämpke, Martin Firlus, Frank Scholz, and Andre Felgner.

This work was jointly funded by the European Metrology Research Programme (EMRP) participating countries within the European Association of National Metrology Institutes (EURAMET) and the European Union.

It was also supported in part by the Grant-in-Aid for Scientific Research (B) (KAKENHI 24360037) from the Japan Society for the Promotion of Science.

Certain commercial equipment, instruments, or materials are identified in this paper in order to specify the experimental procedure adequately. Such identification is not intended to imply recommendation or endorsement by the National Institute of Standards and Technology, nor is it intended to imply that the materials or equipment identified are necessarily the best available for the purpose.